\documentclass[reprint,superscriptaddress,amssymb,amsmath,aps,prl,nobibnotes]{revtex4-2}
\usepackage{soul}
\usepackage[pdftex]{graphicx}
\usepackage{amsmath}
\usepackage{color}
\usepackage{natbib}
\usepackage{xcolor}

\begin{document}

\title{Complexity in the medium-range order of gallium as a polyvalent liquid metal}


\author{Chengyun Hua}

\affiliation{Materials Science and Technology Division, Oak Ridge National Laboratory, Oak Ridge, Tennessee, 37831 USA}

\author{Yadu K. Sarathchandran}
\affiliation{Department of Physics and Astronomy, The University of Tennessee, Knoxville, Tennessee, 37996, USA}

\author{Eva Zarkadoula}
\affiliation{Center for Nanophase Materials Sciences, Oak Ridge National Laboratory, Oak Ridge, Tennessee, 37831 USA}

\author{Wojciech Dmowski}
\affiliation{Department of Materials Science and Engineering, The University of Tennessee, Knoxville, Tennessee, 37996 USA}

\author{Douglas L. Abernathy}
\affiliation{Neutron Scattering Division, Oak Ridge National Laboratory, Oak Ridge, Tennessee, 37831 USA}

\author{Yuya Shinohara}
\affiliation{Materials Science and Technology Division, Oak Ridge National Laboratory, Oak Ridge, Tennessee, 37831 USA}

\author{Takeshi Egami}

\affiliation{Materials Science and Technology Division, Oak Ridge National Laboratory, Oak Ridge, Tennessee, 37831 USA}
\affiliation{Department of Physics and Astronomy, The University of Tennessee, Knoxville, Tennessee, 37996, USA}
\affiliation{Department of Materials Science and Engineering, The University of Tennessee, Knoxville, Tennessee, 37996 USA}

\date{\today}

\begin{abstract}

Simplicity in chemical composition does not always translate into simplicity in the structures and dynamics of liquids and solids. Some elementary liquid metals, such as gallium, show unusual behaviors in thermodynamic and transport properties as a result of their complex atomic structure and dynamics. In this work, we study the real-space atomic correlation function of liquid gallium by neutron scattering. In the pair-distribution function, there exist two kinds of medium-range order (MRO), characterized by oscillations beyond the first nearest neighbors. On the other hand, the first neighbor shell shows only one kind of bond. The two types of MRO are strongly overlapping in space and fluctuating in time. We propose that they are the basis for anomalous behavior of liquid gallium. This view challenges the current view that liquid gallium consists of fluctuating metallic and insulating domains. These findings shed new light on the interpretation of similar microscopic anomalies observed in other semi-metallic liquids. 
\end{abstract}
\maketitle

In the periodic table, elements near the left edge of the periodic table are typically metals, characterized by their tendency to lose electrons and form metallic bonds, while those near the right edge are either chemically inert (such as the noble gases) or exhibit strong electron affinity, often forming ionic or covalent compounds. Those in the middle, such as Ga, Si, Ge, Sn, and Bi, are polyvalent elements and exhibit complex thermodynamic and transport properties \cite{waseda_structure_1972, hafner_structural_1992, kawakita_structure_2005, kawakita_anomaly_2018}. These elements usually have low melting temperatures and are more conductive in their liquid phase than in the solid phase. Gallium even shows a higher density in the liquid state than in the solid state, similar to the behaviors of water, which is an atypical liquid, partly due to hydrogen bonding \cite{finney_structure_2024}, and exhibits an anomalous change in diffusivity as a function of temperature \cite{balyakin_viscosity_2022, cornell_structure_1967, petit_self-diffusion_1956,gong_gallium_1991}. The atomic structure and collective excitation in the solid state of these polyvalent elements have been widely studied, but little is known about these elements in the liquid state.  

One well-known characteristic of certain polyvalent elements in the liquid state---such as gallium, bismuth, lead, and silicon---is the presence of a structural anomaly in their structure factor, $S(Q)$, where $Q$ is the momentum exchange of scattering. This anomaly manifests as a distinct shoulder adjacent to the first peak. Many previous studies have attributed this asymmetric feature at $Q_{\mathrm{shoulder}}$ in reciprocal space \cite{gong_coexistence_1993,nield_changes_1998,lambie_resolving_2024,yang_firstprinciples_2011} to coexistence of metallic and non-metallic local environments, emphasizing short-range order (SRO) in the form of locally ordered metallic or covalent structures \cite{mokshin_short-range_2015}. However, this interpretation remains controversial, as no direct evidence for the existence of a mixed bonding environment has been demonstrated \cite{lambie_resolving_2024,nield_changes_1998,gong_coexistence_1993,yang_firstprinciples_2011,tsai_revisiting_2010,blagoveshchenskii_microscopic_2014,mokshin_short-range_2015,shahzad_atomic_2024}.

In this study, using liquid gallium as an example, we demonstrate that the first peak of $S(Q)$ is directly linked to medium-range order (MRO) rather than SRO. Here, MRO refers to structural correlations that extend beyond the nearest neighbor shell \cite{hansen_theory_2013}, manifesting over length scales of several atomic diameters, and is crucial to determine the transport and mechanical properties of liquids \cite{shi_revealing_2023,sorensen_revealing_2020}. Recent studies \cite{iwashita_seeing_2017, egami_correlated_2020} suggest that these intermediate-scale correlations arise from density fluctuations driven by collective atomic arrangements. Thus we propose here that the anomalous feature near the first peak of $S(Q)$ reflects the inherent complexity of MRO in polyvalent liquids.

\begin{figure*}[t]
    \includegraphics[scale = 0.35]{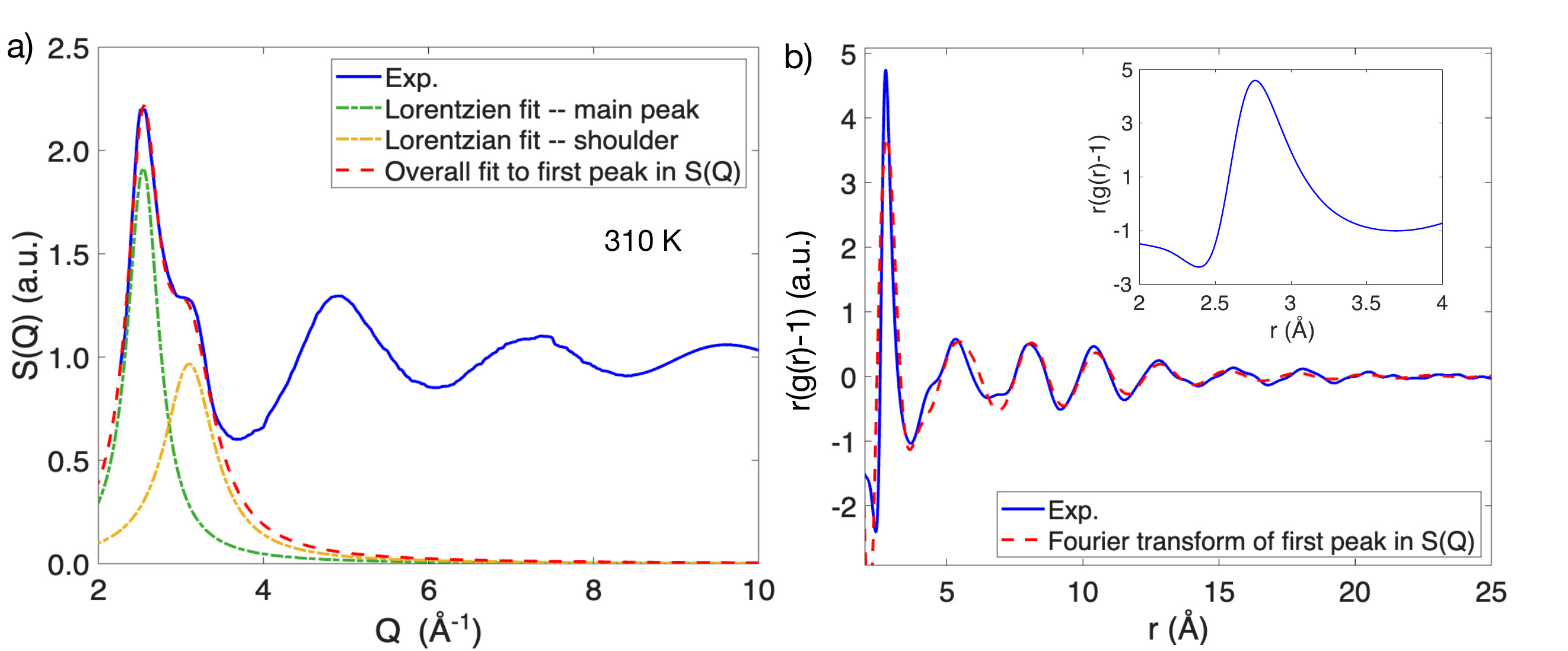}
    \caption{ (a) The measured structure factor, $S(Q)$ at 310 K (solid line), two superimposed Lorentzian fit to the first peak of $S(Q)$ (dashed line), and the decomposed Lorentzian functions describing the main peak and the shoulder (dotted-dashed line).  (b) The measured pair distribution function, $r(g(r)-1)$ (solid line) and the Fourier transformation of the fit to the first peak of $S(Q)$  (dashed line). Inset: Zoom-in view of the first peak in $g(r)$.}
    \label{fig:StaticSF}
\end{figure*}

We study the real-space atomic correlation function of liquid gallium by neutron scattering. The snapshot structure factor, $S(Q)$, shown in Fig.~\ref{fig:StaticSF}(a), was obtained by integrating the dynamic structure factor, $S(Q,E)$, over the energy transfer $E$, as measured by inelastic neutron scattering on the wide-Angular Range Chopper Spectrometer (ARCS) \cite{abernathy_ARCS_2012} at the Spallation Neutron Source (SNS), Oak Ridge National Laboratory. A detailed account on the experimental method and data reduction is provided elsewhere \cite{hua_prb_2025}, and a brief description is included in the Supplemental Material \cite{supp}. The real-space atomic correlation function or the atomic pair-distribution function (PDF), $g(r)$, is then obtained as an inverse Fourier transform of $S(Q)$. Figure~\ref{fig:StaticSF}(b) shows $g(r)$ of liquid gallium at 310~K presented as $r\left[g(r)-1\right]$. 

In $S(Q)$, the first positive correlation at 2.55\,$\mathrm{\AA}^{-1}$ has a shoulder on the high-$Q$ side of the peak at $Q_{\mathrm{shoulder}} = 3.1 \,\mathrm{\AA}^{-1}$, in good agreement with previous X-ray and neutron diffraction studies \cite{narten_liquid_1972}. The shoulder, or asymmetric feature, at $Q_{\mathrm{shoulder}}$ in reciprocal space has attracted attention of researchers, but its interpretation in real space structure has been controversial. In what follows, we propose that MRO is the origin of the asymmetric feature at $Q_{\mathrm{shoulder}}$ and reflects the nature of the structural state in polyvalent liquid metals. 

\begin{figure}[t]
    \includegraphics[scale = 0.4]{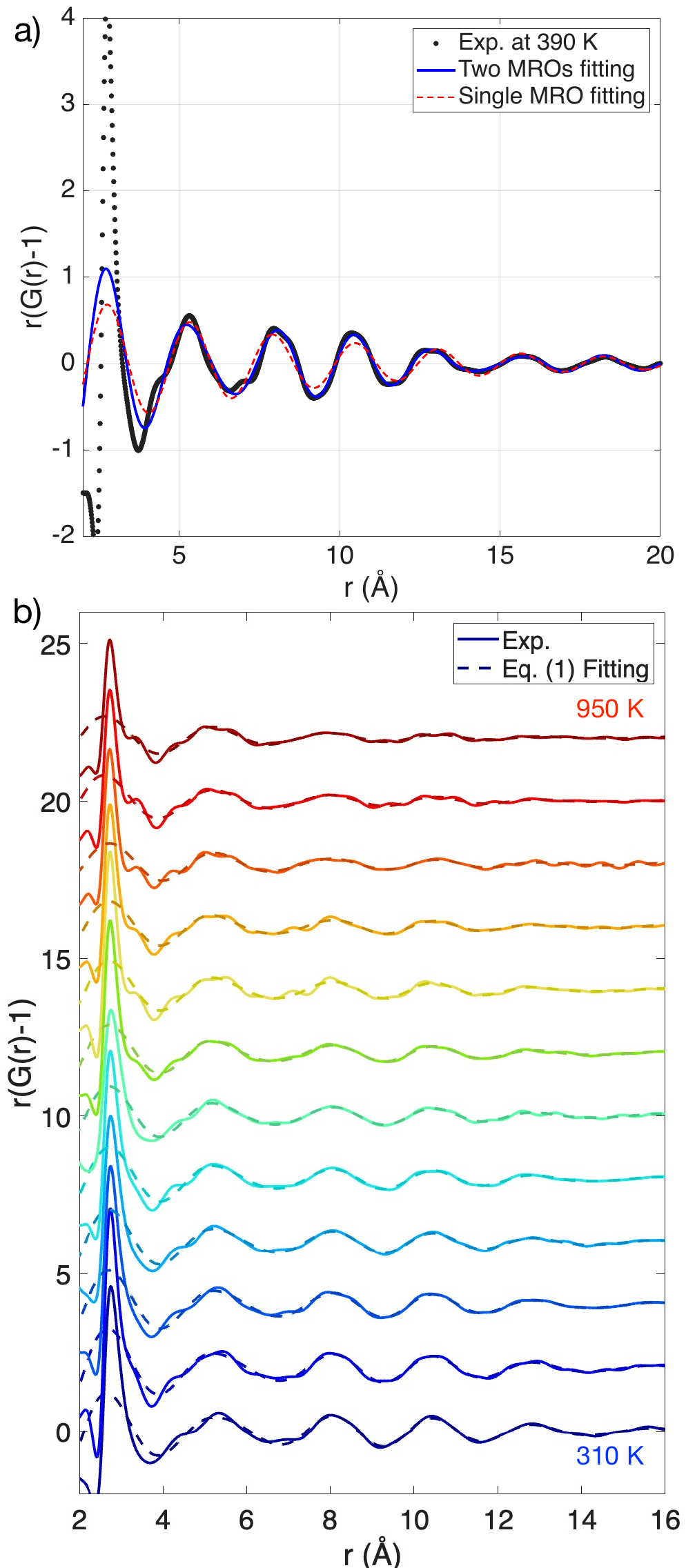}
    \caption{(a) The measured pair distribution function, $r(g(r)-1)$ (dotted line) and the best fit of the higher-order peaks ($r > 3.8 \mathrm{\AA}$) using Eq.~(\ref{eq:MRO}) with one (dashed line) and two (solid line) sinusoidal components at 390 K. (b) The full set of the measured $r(g(r)-1)$ with the best fit using Eq. ~(\ref{eq:MRO}) between 310 K (bottom) and 950 K (top). The values on the y-axis correspond to the curve at 310 K; the remaining solid curves are displaced upwards with a step of 2. The full set of temperatures is: 310, 350, 390, 430, 470, 510, 550, 650, 750, 850, and 950 K.}
    \label{fig:twoMROs}
\end{figure}

Because $S(Q)$ and $g(r)$ are connected by the Fourier transformation, the first peak of $S(Q)$ at $Q_1$ is often assumed to correspond to the first peak of $g(r)$ at $r_1$ by $r_1 = 2\pi/Q_1$. However, this assumption is incorrect \cite{cargill_structure_1975}. Indeed, while the first positive correlation of $S(Q)$ has a prominent shoulder, the first neighbor correlation at around $r = 2.8\,\mathrm{\AA}$ does not have a shoulder as shown in Fig.~\ref{fig:StaticSF}(b). If multiple types of nearest-neighbor shell configurations are present, or if mixed molecular clusters with distinct structural characteristics form, the first peak of $g(r)$ would likely be split---or at least appear asymmetric. For instance, \emph{ab initio} calculations of liquid gallium predict that the bond length scales corresponding to covalent and metallic coordination shells are $\lesssim$2.47 $\mathrm{\AA}$ and $\sim$2.7--2.8 $\mathrm{\AA}$, respectively \cite{lambie_resolving_2024}. However, the $g(r)$ of liquid gallium at 310 K in Fig.~\ref{fig:StaticSF}(b) has a single, well-defined first peak, meaning only one structural environment is observable in the measurement. This difference between the first peaks of $S(Q)$ and $g(r)$ suggests that they do not correspond to the same structural feature, but instead reflect distinct structural regimes in liquids.

To describe the first peak in $S(Q)$, we fit two Lorentzian functions (dashed line in Fig.~\ref{fig:StaticSF}(a)) to $S(Q)$ up to 3.8 $\mathrm{\AA}$. The fitted first peak is then Fourier transformed to the real space (dashed line in Fig.~\ref{fig:StaticSF}(b)). It is clear that the extended MRO represented by the oscillation well beyond 10\,\AA~in $g(r)$ is directly linked to the first peak in $S(Q)$. The first peak in $S(Q)$ reflects the higher-order correlations in $g(r)$, thus MRO, whereas the first peak of $g(r)$ is primarily influenced by the higher-order features of $S(Q)$ \cite{egami_origin_2023}. For a simple liquid, the long-distance oscillations in $[g(r)-1]$ are approximately described by an exponentially-decaying sinusoidal function in the form of $\sin(rQ_{\mathrm{MRO}}+\delta)\exp(-r/\xi_s)/r$ \cite{egami_medium-range_2023, ryu_curie-weiss_2019}. This functional form arises from the Fourier transformation of a shifted Lorentzian function, where $\xi_s$ is the structural coherence length in real space and its inverse value describes the width of the Lorentzian function in reciprocal space. $Q_{\mathrm{MRO}}$ represents the wavelength of the MRO oscillations. $Q_{\mathrm{MRO}}$ is very close to the position of the first peak of $S(Q)$, $Q_1$. For gallium,  because the first peak and its shoulder of $S(Q)$ are described as a superposition of two shifted Lorentzian functions, the oscillatory terms in real space is then given by
\begin{eqnarray}\label{eq:MRO}
 && G_{\mathrm{MRO}}(r)-1  \approx \left[A_{1}\sin(rQ_{\mathrm{MRO}1}+\delta_{1}) \right. \\ 
&& \left.+A_{2}\sin(rQ_{\mathrm{MRO}2}+\delta_{2})\right]\frac{\exp\left(-r/\xi_s\right)}{r}, \ r>r_{\mathrm{cutoff}}  \nonumber
\end{eqnarray}
where $r_{\mathrm{cutoff}} = 3.8~\mathrm{\AA}$ is the position of the first minimum of $g(r)$ beyond the first peak. It is important to emphasize that the details of MROs are unambiguously accessible through real-space correlation functions, \emph{i.e.}, via Eq.~(\ref{eq:MRO}). Although the snapshot structure factor, $S(Q)$, is more readily accessible in experiments and the first peak in $S(Q)$ primarily reflects correlations beyond the first-neighbor shell in $g(r)$, it is still influenced by nearest neighbor correlations. This influence explains the discrepancy observed in the second peak of $g(r)$ obtained by a direct Fourier transform of first peak in $S(Q)$. To decompose the two superimposed oscillations and extract parameters such as the amplitude and wavelength of each, it is necessary to use Eq.~(\ref{eq:MRO}) to analyze the extended oscillations in $g(r)$, not in $S(Q)$. Figure~\ref{fig:twoMROs}(a) compares the fitting results to the measured $g(r)$ at 390 K using Eq.~(\ref{eq:MRO}) with one and two  sinusoidal components. For gallium, a single sinusoidal function provides a poor fit to the measured $g(r)$, whereas incorporating two MROs more accurately captures the oscillatory behavior. 

The best fits to $g(r)$ using Eq.~(\ref{eq:MRO}) at all the temperature range we measured are shown as dashed lines in Fig.~\ref{fig:twoMROs}(b). Three major features emerge from the analysis: (1) the amplitudes of the two sinusoidal components are of the same order of magnitude, \emph{i.e.}, $A_1 = 1.82$ and $A_2 = 0.75$ at 310 K, indicating that both are essential to accurately capture the MRO oscillations; (2) a single coherence length, \emph{i.e.}, $\xi_s = 5.88\,\mathrm{\AA}$ at 310 K, adequately characterizes the decay behavior of both oscillations; and (3) the wavevectors closely correspond to the positions of the principal peak and shoulder in $S(Q)$. In the SI, we show that the experimentally measured $S(Q)$ is reproduced with high accuracy by combining the first peak of $g(r)$ and the fitted $g_{\mathrm{MRO}}(r)$ with $r>r_{\mathrm{cutoff}}$ and computing its Fourier transform. Notably, the shoulder feature in $S(Q)$ cannot be reproduced using a single sinusoidal function; instead, two sinusoidal functions are required. This underscores that the shoulder feature in $S(Q)$ is directly associated with higher-order peaks in $g(r)$ and thus originates from MRO, rather than the SRO in the nearest neighbors.

To identify the origin of the two observed MROs, it is essential to distinguish MRO from SRO. While SRO reflects local chemical environments, MRO, as defined by Eq.~(\ref{eq:MRO}), captures coarse-grained density fluctuations on the scale of several Ångströms \cite{Ryu_compositional_2025}. These extended oscillations indicate undulations in atomic density \cite{egami_structural_2022, ryu_medium-range_2021, egami_origin_2023}. The MRO described here thus refers to the strength of such density waves, characterizing a higher, more collective level of structural organization, rather than the fine details of atomic configurations \cite{egami_medium-range_2023}.

Temperature-dependent positions ($Q_{MRO1}$ and $Q_{MRO2}$) and amplitudes ($A_1$ and $A_2$) of the two MROs are provided in Ref.~\cite{hua_prb_2025}.  Both the amplitudes and positions exhibit only weak temperature dependence, suggesting that the two MRO components not only coexist but also persist up to elevated temperatures. The decrease in the oscillation amplitude---or the amplitude of the first peak in $S(Q)$---is attributed to the reduction in structural coherent length as temperature increases. 

The value of $Q_{\mathrm{MRO}2} (= 3.22~\mathrm{\AA}^{-1})$ is very close to $2k_F (= 3.25~\mathrm{\AA}^{-1})$, suggesting the link between the second MRO and electrons. In the dense random packing structure, the MRO periodicity is related to the density by $Q_{\mathrm{MRO}}=C_{\mathrm{MRO}} \rho_0^{1/3}$ , where $C_{\mathrm{MRO}} = 7.04$ and $\rho_0$ is the atomic number density \cite{egami_origin_2023}. This yields $Q_{\mathrm{MRO}} = 2.64~\mathrm{\AA}^{-1}$, which is close to $Q_{\mathrm{MRO}1} (= 2.47~\mathrm{\AA}^{-1})$. The small difference may be accounted for by the fact that gallium is not close-packed, with an average coordination of $\sim$10.5 at 310 K \cite{li_local_2017}, whereas in the dense random packing structure, the coordination number is expected to be approximately $4\pi$ (12.56) \cite{egami-aur_1987}. Therefore, we surmise that MRO1 is driven by the system's tendency to form a random-packed structure, whereas MRO2 is driven by electrons to form charge density waves.

It is worth mentioning that Friedel oscillations and MRO2 originate from the same underlying cause---the electronic structure of gallium metal. Indeed, many previous studies have suggested the role of Friedel oscillations \cite{tsai_revisiting_2010, tsai_entropy_2011,hafner_structural_1990, hafner_structural_1992, jank_structural_1990-1,jank_structural_1990} in reproducing the shoulder feature in $S(Q)$. However, there is a crucial distinction between them: Friedel oscillations decay as $1/r^3$ with distance, whereas MRO oscillations decay as $1/r$, making them far more extended. We speculate that this difference arises from the distinct nature of the underlying phenomena—Friedel oscillations represent a local screening response to a point charge, while MRO2 oscillations correspond to a collective density-wave response to many charge centers randomly distributed in space. Thus, MRO2 is not simply a direct consequence of Friedel contributions to the interatomic potential, as is often assumed, but rather a more collective electronic response better understood in terms of the virtual formation of charge-density waves.

\begin{figure}[t]
    \includegraphics[scale = 0.45]{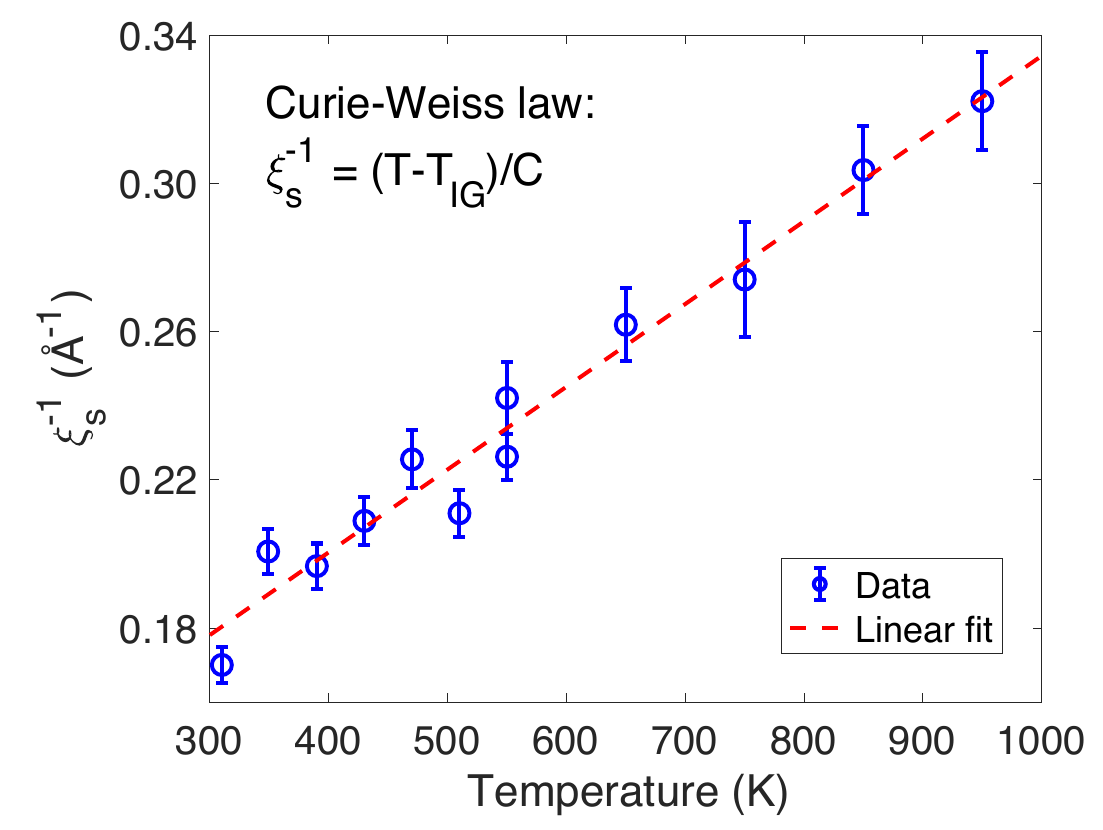}
    \caption{ The temperature dependence of inverse coherent length, $\xi_s^{-1}$. $\xi_s(T)$ follows a Curie-Weiss law with $T_{\mathrm{IG}} = -498~\mathrm{K}$.). 
    }
    \label{fig:CW}
\end{figure}

Despite their different microscopic origins, the two types of MRO in liquid gallium shares a single coherent length, indicating a tendency to form overlapping, idealized coherent states. In Fig.~\ref{fig:CW}, we show that the relationship between the structural coherence length and temperature strictly follows the Curie-Weiss law \cite{ryu_curie-weiss_2019}, $\xi_s^{-1}(T) \propto T - T_{\mathrm{IG}}$. This law suggests a temperature-dependent increase in $\xi_s(T)$, implying a tendency toward the formation of a long-range density wave state as temperature decreases \cite{egami_origin_2023}. In the case of gallium, such a hypothetical state would occur at $T_{\mathrm{IG}} = -498$ K. Although this state is not physically realizable—both due to the negative value of $T_{\mathrm{IG}}$ and the freezing of $\xi_s(T)$ at the glass transition \cite{ryu_ideality_2020}—the emergence of Curie-Weiss behavior still indicates the presence of an underlying thermodynamic driving force toward long-range structural order in liquid gallium. Deviations from these ideal states arise primarily from thermal density fluctuations at elevated temperatures \cite{egami_medium-range_2021}. As a result, both MROs exhibit the same spatial decay behavior, characterized by a common structural coherence length.

In addition to their shared spatial characteristics, in Ref.~\cite{hua_prb_2025}, we show that the two MROs also display the same temporal decay behavior at all temperatures, where both MRO amplitudes decay with a single exponential decay function, $\exp(-t/\tau_{\mathrm{MRO}})$. This exponential form reflects the stochastic nature of local structural rearrangements—such as bond breaking and formation—which are governed by Poisson statistics. In this framework, $\tau_{\mathrm{MRO}}$ can be interpreted as the expected time interval between successive local structural rearrangements \cite{Bellisard_2018}. At room temperature, $\tau_{\mathrm{MRO}}$ is approximately 2 ps—much longer than time scale of phonon-like excitations ($\sim$ 0.1 ps) \cite{scopigno_high-frequency_2002, demmel_slow_2020,bove_vibrational_2005,khusnutdinoff_collective_2020} or the lifetime of covalent bonds predicted by simulations in  gallium \cite{lambie_resolving_2024}. This significant difference in timescales emphasizes that MRO dynamics represent collective behavior involving many atoms.

The idea that two types of strongly overlapping and rapidly fluctuating MROs underlie the anomalous behavior of liquid gallium offers a perspective fundamentally different from the prevailing view that liquid gallium consists of fluctuating metallic and insulating domains. In fact, the coexistence of two types of MRO may be more common than previously recognized. In metals such as Mg and Ca, for which valence electron-per-atom ratio is equal to 2, the principal peak position, $Q_1$ in $S(Q)$ is approximately equal to $2 k_F$, and the long-range oscillations induced by the electronic structure are not in conflict with the requirement of local packing \cite{hafner_atomic_1988}. However, in metals and metalloids from groups IIIA, IVA, VA, and VIA, the valence-electron-per-atom ratio is greater than 2, resulting in $2 k_F > Q_p$. In such cases, local packing that promotes the MRO at $Q_{\mathrm{MRO}1}$ and the electronic drive that favors the MRO at $Q_{\mathrm{MRO}2}$ are in competition, leading to the formation of two distinct MROs. Whereas this idea still needs further study in metal melts other than gallium, it provides a basis for developing a more realistic and general theory in liquid metals that have shown structural anomalies. More importantly, the new insight based on the density wave approach offers a promising research direction to connect the electronic structure to atomic dynamics and understand the dynamic nature of liquid and glass.

This work was supported by U.S. Department of Energy (DOE), Office of Science, Office of Basic Energy Science (BES), Division of Materials Sciences and Engineering. A portion of this research used the resources at the Spallation Neutron Source, supported by DOE, BES, Scientific User Facilities Division. The beamtime was allocated to ARCS on proposal number IPTS-21883. This research used resources of the National Energy Research Scientific Computing Center (NERSC), a Department of Energy User Facility using NERSC award BES-ERCAP0031522.

\bibliography{MyRef.bib}

\end{document}